\documentclass
[pra,onecolumn,letterpaper,noshowpacs,preprint,12pt,titlepage]{revtex4}%
\usepackage{amsfonts}
\usepackage{amsmath}
\usepackage{amssymb}
\usepackage{graphicx}%
\begin{document}
\title{Conditions for the quantum adiabatic approximation}
\author{Ming-Yong Ye}
\author{Xiang-Fa Zhou}
\author{Yong-Sheng Zhang}
\email{yshzhang@ustc.edu.cn}
\author{Guang-Can Guo}
\email{gcguo@ustc.edu.cn}
\affiliation{Key Laboratory of Quantum Information, Department of Physics, University of
Science and Technology of China (CAS), Hefei 230026, People's Republic of China}

\begin{abstract}
We give a sufficient condition for the quantum adiabatic approximation, which
is quantitative and can be used to estimate error caused by this
approximation. We also discuss when the traditional condition is sufficient.

PACS number(s): 03.65.Ca, 03.65.Ta, 03.65.Vf

\end{abstract}
\maketitle

Quantum adiabatic approximation has a long history but recently it is
re-examined \cite{marzlin,sarandy,wu,tong}. Tong \textit{et al} point out that
the traditional used condition is insufficient, which does not guarantee the
validity of the adiabatic approximation \cite{tong}. In this paper we present
a sufficient condition. Our condition is a quantitative one and can be used to
estimate error caused by the approximation.

When we make an approximation two things should be specified: (i) A time
interval in which we use the approximation and (ii) a parameter that is used
to show how good the approximation is. To our knowledge these two things have
not been specified for adiabatic approximation. Usually quantum adiabatic
approximation is expressed as follows: suppose a quantum system has a finite
and discrete spectral decomposition, if at the start time $t=0$ the system is
in the $n$th instantaneous eigenstate, then at a later time the system will
remain in the $n$th instantaneous eigenstate up to a phase factor provided the
system Hamiltonian varies slowly enough.

Assume the system has finite instantaneous discrete spectrum decomposition%

\[
H(t)=\sum_{m}E_{m}(t)\left\vert E_{m}(t)\right\rangle \left\langle
E_{m}(t)\right\vert .
\]
And the initial state of the system is
\[
\left\vert \Psi\left(  0\right)  \right\rangle =%
{\displaystyle\sum\limits_{n}}
c_{n}\left(  0\right)  \left\vert E_{n}(0)\right\rangle .
\]
The adiabatic approximation predicates that at time $t$ the system will be in
the state%
\[
\left\vert \Psi_{A}\left(  t\right)  \right\rangle =%
{\displaystyle\sum\limits_{n}}
c_{n}\left(  0\right)  e^{-i\alpha_{n}\left(  t\right)  }e^{-i\beta_{n}\left(
t\right)  }\left\vert E_{n}(t)\right\rangle ,
\]
Where $\alpha_{n}\left(  t\right)  =\int_{0}^{t}E_{n}(t^{\prime})dt^{\prime}$
is the dynamic phase, and $\beta_{n}\left(  t\right)  =-i\int_{0}^{t}\left(
\left\langle E_{n}(t^{\prime})\right\vert \frac{d}{dt^{\prime}}\left\vert
E_{n}(t^{\prime})\right\rangle \right)  dt^{\prime}$ is Berry phase
\cite{berry}. We specify a time interval $[0,T]$ in which we use the adiabatic
approximate evolution to simulate the real evolution of the system. Our goal
is to find the condition under which the error caused by the approximation is
below a value $\epsilon$ we specified.

Suppose the true wave function at time $t$ is
\[
\left\vert \Psi\left(  t\right)  \right\rangle =%
{\displaystyle\sum\limits_{n}}
c_{n}\left(  t\right)  e^{-i\alpha_{n}\left(  t\right)  }e^{-i\beta_{n}\left(
t\right)  }\left\vert E_{n}(t)\right\rangle .
\]
This wave function obeys the Schr\"{o}dinger equation $id\left\vert
\Psi\left(  t\right)  \right\rangle /dt=H(t)\left\vert \Psi\left(  t\right)
\right\rangle $ (we set $\hbar=1$). From $i\left\langle E_{m}(t)\right\vert
d\left\vert \Psi\left(  t\right)  \right\rangle /dt=\left\langle
E_{m}(t)\right\vert H(t)\left\vert \Psi\left(  t\right)  \right\rangle $ we
obtain%
\[
\frac{dc_{m}\left(  t\right)  }{dt}=-\sum_{n\neq m}c_{n}\left(  t\right)
\left\langle E_{m}\right\vert \frac{d}{dt}\left\vert E_{n}\right\rangle
e^{-i\left(  \alpha_{n}\left(  t\right)  -\alpha_{m}\left(  t\right)  \right)
}e^{-i\left(  \beta_{n}\left(  t\right)  -\beta_{m}\left(  t\right)  \right)
}.
\]
This differential equation is equivalent to the following integral equation%
\[
c_{m}\left(  t\right)  -c_{m}\left(  0\right)  =-\sum_{n\neq m}\int_{0}%
^{t}c_{n}\left(  t^{\prime}\right)  \left\langle E_{m}\right\vert \frac
{d}{dt^{\prime}}\left\vert E_{n}\right\rangle e^{-i\left(  \alpha_{n}\left(
t^{\prime}\right)  -\alpha_{m}\left(  t^{\prime}\right)  \right)
}e^{-i\left(  \beta_{n}\left(  t^{\prime}\right)  -\beta_{m}\left(  t^{\prime
}\right)  \right)  }dt^{\prime}.
\]
We assume the adiabatic approximation is very good in the time interval
$[0,T]$, which means $\left\vert c_{m}\left(  t\right)  -c_{m}\left(
0\right)  \right\vert \ll1$ for arbitrary initial value $c_{m}\left(
0\right)  $. Mathematically this assumption requires%
\begin{equation}
\left\vert \int_{0}^{t}\left\langle E_{m}\right\vert \frac{d}{dt^{\prime}%
}\left\vert E_{n}\right\rangle e^{-i\left(  \alpha_{n}\left(  t^{\prime
}\right)  -\alpha_{m}\left(  t^{\prime}\right)  \right)  }e^{-i\left(
\beta_{n}\left(  t^{\prime}\right)  -\beta_{m}\left(  t^{\prime}\right)
\right)  }dt^{\prime}\right\vert \ll1,t\in\lbrack0,T]\label{1}%
\end{equation}
for any $n\neq m.$ Recall a simple physical picture where a two-level atom is
presented in a classical field. When the coupling strength between the atom
and the field is much smaller than the energy detuning, the population of the
atom will not change. Compare the question we are considering with this simple
physical picture, intuitively we conjecture that when the adiabatic
approximation is a good approximation in the time interval $[0,T]$\ we should
have%
\begin{equation}
\left\vert \frac{\left\langle E_{m}\right\vert \frac{d}{dt^{\prime}}\left\vert
E_{n}\right\rangle }{\frac{d}{dt^{\prime}}\left[  \alpha_{n}\left(  t^{\prime
}\right)  -\alpha_{m}\left(  t^{\prime}\right)  +\beta_{n}\left(  t^{\prime
}\right)  -\beta_{m}\left(  t^{\prime}\right)  +\gamma_{mn}\left(  t^{\prime
}\right)  \right]  }\right\vert \ll1,t^{\prime}\in\lbrack0,T]\label{2}%
\end{equation}
for any $m\neq n$, where $\gamma_{mn}\left(  t^{\prime}\right)  $ is the phase
of $\left\langle E_{m}\right\vert \frac{d}{dt^{\prime}}\left\vert
E_{n}\right\rangle $, \textit{i.e.}, $\left\langle E_{m}\right\vert \frac
{d}{dt^{\prime}}\left\vert E_{n}\right\rangle =\left\vert A_{mn}\left(
t^{\prime}\right)  \right\vert e^{-i\gamma_{mn}\left(  t^{\prime}\right)  }$.
We will show that this requirement can be regarded as a qualitative condition
for quantum adiabatic approximation.

In the above we have specified the time interval $[0,T]$\ we use the adiabatic
approximation. Now we specify a small number $\epsilon$ to show how good the
approximation will be in this time interval. We require%
\begin{equation}
\left\vert \int_{0}^{t}\left\langle E_{m}\right\vert \frac{d}{dt^{\prime}%
}\left\vert E_{n}\right\rangle e^{-i\left(  \alpha_{n}\left(  t^{\prime
}\right)  -\alpha_{m}\left(  t^{\prime}\right)  \right)  }e^{-i\left(
\beta_{n}\left(  t^{\prime}\right)  -\beta_{m}\left(  t^{\prime}\right)
\right)  }dt^{\prime}\right\vert \leq\epsilon,t\in\lbrack0,T]\label{3}%
\end{equation}
for any $n\neq m.$ Obviously the smaller $\epsilon$ is, the better the
adiabatic approximation will be. So the parameter $\epsilon$ can be used to
tell how good the adiabatic approximation is in the time interval $[0,T]$. We
think this inequality is the sufficient and necessary condition for quantum
adiabatic approximation. When we want to know whether quantum adiabatic
approximation is a good approximation (specified by $\epsilon$ ) for a system
evolution in the time interval $[0,T]$, we can check whether the system
Hamiltonian satisfies the above inequality or not. When this check is not easy
to accomplish, some sufficient conditions that are easy to check can be used.

We define%
\begin{align*}
G_{mn} &  =\left\vert \int_{0}^{t}\left\langle E_{m}\right\vert \frac
{d}{dt^{\prime}}\left\vert E_{n}\right\rangle e^{-i\left(  \alpha_{n}\left(
t^{\prime}\right)  -\alpha_{m}\left(  t^{\prime}\right)  \right)
}e^{-i\left(  \beta_{n}\left(  t^{\prime}\right)  -\beta_{m}\left(  t^{\prime
}\right)  \right)  }dt^{\prime}\right\vert \\
&  =\left\vert \int_{0}^{t}\left\vert \left\langle E_{m}\right\vert \frac
{d}{dt^{\prime}}\left\vert E_{n}\right\rangle \right\vert e^{-i\theta
_{mn}\left(  t^{\prime}\right)  }dt^{\prime}\right\vert ,
\end{align*}
where $\theta_{mn}\left(  t^{\prime}\right)  =\alpha_{n}\left(  t^{\prime
}\right)  -\alpha_{m}\left(  t^{\prime}\right)  +\beta_{n}\left(  t^{\prime
}\right)  -\beta_{m}\left(  t^{\prime}\right)  +\gamma_{mn}\left(  t^{\prime
}\right)  $. Notice that
\[
G_{mn}\leq\int_{0}^{t}\left\vert \left\langle E_{m}\right\vert \frac
{d}{dt^{\prime}}\left\vert E_{n}\right\rangle \right\vert dt^{\prime}\leq
T\max_{t^{\prime}\in\lbrack0,T]}\left\vert \left\langle E_{m}\right\vert
\frac{d}{dt^{\prime}}\left\vert E_{n}\right\rangle \right\vert .
\]
So when
\begin{equation}
T\max_{t^{\prime}\in\lbrack0,T]}\left\vert \left\langle E_{m}\right\vert
\frac{d}{dt^{\prime}}\left\vert E_{n}\right\rangle \right\vert \leq
\epsilon\label{4}%
\end{equation}
for any $m\neq n$ we can say in the time interval $[0,T]$\ the adiabatic
approximation is a good approximation under the error rate we specify by
$\epsilon$. Physically this situation means the eigenstates change little in
the time interval $[0,T]$. In the limit case, \textit{i.e.}, $\left\langle
E_{m}\right\vert \frac{d}{dt^{\prime}}\left\vert E_{n}\right\rangle =0$ for
any $m\neq n$, $\frac{d}{dt^{\prime}}\left\vert E_{n}\right\rangle $ is always
proportional to $\left\vert E_{n}\right\rangle $ or $\left\vert E_{n}%
\right\rangle $ is a constant vector$,$ which means $\left\vert E_{n}\left(
t^{\prime}\right)  \right\rangle $ is equivalent to $\left\vert E_{n}%
(0)\right\rangle $ up to a phase factor. When $T\max_{t^{\prime}\in
\lbrack0,T]}\left\vert \left\langle E_{m}\right\vert \frac{d}{dt^{\prime}%
}\left\vert E_{n}\right\rangle \right\vert \leq\epsilon$ is not satisfied, we
make some assumptions for further discussion. (i) We assume $\theta
_{mn}\left(  t^{\prime}\right)  $ is an increasing (or decreasing) function in
the interval $t^{\prime}\in\lbrack0,T]$, so we can write%
\begin{align*}
G_{mn} &  =\left\vert \int_{0}^{t}\left\vert A_{mn}\left(  t^{\prime}\right)
\right\vert e^{-i\theta_{mn}\left(  t^{\prime}\right)  }dt^{\prime}\right\vert
\\
&  =\left\vert \int_{\theta_{mn}\left(  0\right)  }^{\theta_{mn}\left(
t\right)  }\left\vert A_{mn}\left(  t^{\prime}\left(  \theta_{mn}\right)
\right)  \frac{dt^{\prime}\left(  \theta_{mn}\right)  }{d\theta_{mn}%
}\right\vert e^{-i\theta_{mn}}d\theta_{mn}\right\vert \\
&  \leq\left\vert \int_{\theta_{mn}\left(  0\right)  }^{\theta_{mn}\left(
t\right)  }\left\vert A_{mn}\left(  t^{\prime}\left(  \theta_{mn}\right)
\right)  \frac{dt^{\prime}\left(  \theta_{mn}\right)  }{d\theta_{mn}%
}\right\vert \cos\theta_{mn}d\theta_{mn}\right\vert \\
&  +\left\vert \int_{\theta_{mn}\left(  0\right)  }^{\theta_{mn}\left(
t\right)  }\left\vert A_{mn}\left(  t^{\prime}\left(  \theta_{mn}\right)
\right)  \frac{dt^{\prime}\left(  \theta_{mn}\right)  }{d\theta_{mn}%
}\right\vert \sin\theta_{mn}d\theta_{mn}\right\vert .
\end{align*}
(ii) We assume $\left\vert A_{mn}\left(  t^{\prime}\left(  \theta_{mn}\right)
\right)  \frac{dt^{\prime}\left(  \theta_{mn}\right)  }{d\theta_{mn}%
}\right\vert $ is an not-decreasing (or not-increasing) function in the
interval $\theta_{mn}\in\lbrack\theta_{mn}\left(  0\right)  ,\theta
_{mn}\left(  T\right)  ]$, \textit{i.e.}, $\left\vert A_{mn}\left(  t^{\prime
}\right)  \frac{dt^{\prime}}{d\theta_{mn}\left(  t^{\prime}\right)
}\right\vert $ is an not-decreasing (or not-increasing) function in the
interval $t^{\prime}\in\lbrack0,T]$. According to integral mean value theorems
we have%
\begin{align*}
G_{mn} &  \leq4\max_{\theta_{mn}\in\lbrack\theta_{mn}\left(  0\right)
,\theta_{mn}\left(  T\right)  ]}\left\vert A_{mn}\left(  t^{\prime}\left(
\theta_{mn}\right)  \right)  \frac{dt^{\prime}\left(  \theta_{mn}\right)
}{d\theta_{mn}}\right\vert \\
&  =4\max_{t^{\prime}\in\lbrack0,T]}\left\vert \frac{A_{mn}\left(  t^{\prime
}\right)  }{d\theta_{mn}\left(  t^{\prime}\right)  /dt^{\prime}}\right\vert .
\end{align*}
In deriving this inequality we use the following real function integral mean
value theorems: (1) In the interval $x\in\lbrack a,b]$, $f(x)$ is a
not-increasing and not-negative function and $g(x)$ is integrable, then we
have $\int_{a}^{b}f(x)g(x)dx=f(a)\int_{a}^{c}g(x)dx$, $a<c<b$. (2) In the
interval $x\in\lbrack a,b]$, $f(x)$ is a not-decreasing and not-negative
function and $g(x)$ is integrable, then we have $\int_{a}^{b}%
f(x)g(x)dx=f(b)\int_{c}^{b}g(x)dx$, $a<c<b$. Generally the above two
assumptions are incorrect, but we can always divide the interval $[0,T]$ into
$N_{mn}(T)$ small intervals and in each one the above two assumptions are
correct \cite{explain1}. Now we have
\begin{align*}
G_{mn} &  \leq4N_{mn}(T)\max_{t^{\prime}\in\lbrack0,T]}\left\vert \frac
{A_{mn}\left(  t^{\prime}\right)  }{d\theta_{mn}\left(  t^{\prime}\right)
/dt^{\prime}}\right\vert \\
&  =4N_{mn}(T)\max_{t^{\prime}\in\lbrack0,T]}\left\vert \frac{\left\langle
E_{m}\right\vert \frac{d}{dt^{\prime}}\left\vert E_{n}\right\rangle }{\frac
{d}{dt^{\prime}}\left[  \alpha_{n}\left(  t^{\prime}\right)  -\alpha
_{m}\left(  t^{\prime}\right)  +\beta_{n}\left(  t^{\prime}\right)  -\beta
_{m}\left(  t^{\prime}\right)  +\gamma_{mn}\left(  t^{\prime}\right)  \right]
}\right\vert .
\end{align*}
So when%
\begin{equation}
4N_{mn}(T)\max_{t^{\prime}\in\lbrack0,T]}\left\vert \frac{\left\langle
E_{m}\right\vert \frac{d}{dt^{\prime}}\left\vert E_{n}\right\rangle }{\frac
{d}{dt^{\prime}}\left[  \alpha_{n}\left(  t^{\prime}\right)  -\alpha
_{m}\left(  t^{\prime}\right)  +\beta_{n}\left(  t^{\prime}\right)  -\beta
_{m}\left(  t^{\prime}\right)  +\gamma_{mn}\left(  t^{\prime}\right)  \right]
}\right\vert \leq\epsilon\label{5}%
\end{equation}
for any $m\neq n$ we can claim that in the time interval $[0,T]$ the adiabatic
approximation is a good approximation. But when this inequality is not
satisfied we can not certainly claim the adiabatic approximation is not a good
approximation (specified by $\epsilon$) in the time interval $[0,T]$.
Inequalities (4) and (5) are sufficient conditions for the adiabatic
approximation, they are not necessary. Qualitatively inequality (5) implies
\begin{equation}
\left\vert \frac{\left\langle E_{m}\right\vert \frac{d}{dt^{\prime}}\left\vert
E_{n}\right\rangle }{\frac{d}{dt^{\prime}}\left[  \alpha_{n}\left(  t^{\prime
}\right)  -\alpha_{m}\left(  t^{\prime}\right)  +\beta_{n}\left(  t^{\prime
}\right)  -\beta_{m}\left(  t^{\prime}\right)  +\gamma_{mn}\left(  t^{\prime
}\right)  \right]  }\right\vert \ll1.\label{6}%
\end{equation}
This is just our previous conjecture came from analog with a simple physical picture.

When there exists a special basis and the instantaneous eigenstates vectors
$\left\vert E_{n}\right\rangle $ expressed in this basis are always real,
\textit{i.e.}, the system Hamiltonian in this basis is real, we can find that
$\left\langle E_{m}\right\vert \frac{d}{dt^{\prime}}\left\vert E_{n}%
\right\rangle $ is real and Berry phase $\beta_{m}\left(  t^{\prime}\right)  $
is zero. In this situation we have
\begin{align*}
d\theta_{mn}\left(  t^{\prime}\right)  /dt^{\prime} &  =\frac{d}{dt^{\prime}%
}\left[  \alpha_{n}\left(  t^{\prime}\right)  -\alpha_{m}\left(  t^{\prime
}\right)  \right]  \\
&  =E_{n}\left(  t^{\prime}\right)  -E_{m}\left(  t^{\prime}\right)  .
\end{align*}
Inequalities (5) and (6) can be written as
\[
4N_{mn}(T)\max_{t^{\prime}\in\lbrack0,T]}\left\vert \frac{\left\langle
E_{m}\right\vert \frac{d}{dt^{\prime}}\left\vert E_{n}\right\rangle }%
{E_{n}\left(  t^{\prime}\right)  -E_{m}\left(  t^{\prime}\right)  }\right\vert
\leq\epsilon,
\]
and%
\[
\left\vert \frac{\left\langle E_{m}\right\vert \frac{d}{dt^{\prime}}\left\vert
E_{n}\right\rangle }{E_{n}\left(  t^{\prime}\right)  -E_{m}\left(  t^{\prime
}\right)  }\right\vert \ll1.
\]
This inequality is the traditional sufficient condition for quantum adiabatic
approximation, they are correct when Hamiltonian is a real matrix in a certain basis.

The quantum adiabatic approximation is not perfect except for the trival case
where $\left\langle E_{m}\right\vert \frac{d}{dt^{\prime}}\left\vert
E_{n}\right\rangle =0$ for any $m\neq n$, so applying a parameter to estimate
error is necessary when we use this approximation in a certain time interval.
First we can check the inequality (3). When inequality (3) is not easy to
check, we can check inequalities (4) and (5), they are sufficient conditions
for adiabatic approximation. Qualitatively we can check inequality (6).

In conclusion we give a sufficient condition for quantum adiabatic
approximation, which is quantitative and can be used to analyze error caused
by adiabatic approximation. We point out that the widely used traditional
sufficient condition are correct only when the system Hamiltonian can be
presented by a real matrix, Berry phase is always zero in this situation.

This work was funded by National Fundamental Research Program (Program No.
2001CB309300), National Natural Science Foundation of China (No. 10304017),
and Chinese Innovation Fund (No. Grant 60121503).


\begin{thebibliography}{9}
\bibitem {explain1}Here we assume $\theta_{mn}\left(  t^{\prime}\right)  $ is
not a constant.
\end{thebibliography}
\end{document}